\documentclass[twocolumn,showpacs,prd]{revtex4}
\usepackage{dcolumn}
\usepackage{bm}
\usepackage[normalem]{ulem}
\begin{document}

\title{Erroneous Wave Functions of Ciuchi {\it et  al} for Collective Modes \\   
in Neutron Production on Metallic Hydride Cathodes}
\author{A. Widom}
\affiliation{Physics Department, Northeastern University, Boston MA 02115}
\author{Y.N. Srivastava}
\affiliation{Physics Department \& INFN, University of Perugia, Perugia IT}
\author{L. Larsen}
\affiliation{Lattice Energy LLC, 175 North Harbor Drive, Chicago IL 60601}

\begin{abstract}
There is a recent comment\cite{Ciuchi:2012} concerning the theory of collective many body 
effects on the neutron production rates in a chemical battery cathode. Ciuchi {\it et al} 
employ an inverse beta decay expression that contains a two body 
amplitude. Only one electron and one proton may exist in the Ciuchi {\it et al} model initial 
state wave function. A flaw in their reasoning is that one cannot in reality describe 
collective many body correlations with only a two particle wave function. One needs 
very many particles to describe collective effects. In the model wave functions 
of Ciuchi {\it et al} there are no metallic hydrides, there are no cathodes and there are no 
chemical batteries. Employing a wave function with only one electron and one proton is 
inadequate for describing collective metallic hydride surface quantum plasma physics in 
cathodes accurately.
\end{abstract}


\maketitle

\section{Initial Comments \label{com}}
In years past we have been working on weak interaction inverse beta 
decay while interacting with various collective modes of motion in 
condensed matter systems. Our considerations have been recently 
criticized\cite{Ciuchi:2012}. The difference of opinion on the rate 
of neutron production in hydride battery cathodes has a brief 
history
 starting from a talk at Roma La Sapienza by Y. Srivastava cited by  
Ciuchi {\it et al}\cite{Ciuchi:2012}. 

At this talk, a discussion arose where some of  the authors of \cite{Ciuchi:2012} mentioned 
disagreements with the results presented by YS by factors of $10^{40}$, later reduced to 
a factor of $10^{20}$.
We pointed out that our estimates were based on an actual calculation of the collective process 
and directed them to references \cite{WL1:2006,WL2:2007} {\bf } where inverse beta decay had 
been proposed as a mechanism to activate neutron production. Subsequently,  some members of 
that group calculated the inverse beta decay and in an internal report concluded that  we were still 
$10^7$ high in our estimates of neutron production rates\cite{Polosa}.
Towards the goal of reaching full agreement,  we  suggested that they work in analogy 
to the muon inverse decay process. As a result, \cite{Ciuchi:2012}  the initial 
 disagreement  in neutron production rates between us has presently mowed {\em way down from
 forty to a mere two orders of magnitude}.

Our purpose in this note is to give  in public the last needed corrections to the 
Ciuchi {\it et al} model that would bring the results of their  calculation 
in line with  theoretical results in collective mode studies on this subject\cite{WL1:2006,WL2:2007} 
and most recent experimental\cite{Cirillo:2012} findings. A complete discussion of the issues involved is 
under preparation and will be presented shortly.

\section{Danger in the Numbers \label{DN}}

The Ciuchi {\it et al} team asserts that the factor of two or three orders 
of magnitude would render the inverse beta decay unobservable. 
Fortunately they {\it are completely incorrect} in this regard. There 
are experiments carried out by those in 
D. Cirillo {\it et al}\cite{Cirillo:2012} that reside in Naples. They have 
actually observed both nuclear transmutations and actual neutrons in 
hydride metallic battery cathodes. Even if our theoretical neutron 
counting rates were high by a factor of \begin{math} 300 \end{math}, 
then Cirillo {\it et al} could still and indeed did experimentally observe nuclear 
transmutations.

Ciuchi et. al. use our numbers from papers dealing with other applications
but {\it not} batteries. For example, they start 
from neutron production rate with the time honored formula 
\begin{equation}
\Gamma (e^- p^+ \to n + \nu_e )=|\psi (0)|^2 v \sigma 
\label{Ciuchi1}
\end{equation}
wherein the amplitude for finding one electron at position 
\begin{math} {\bf r}  \end{math} and one proton at position 
\begin{math} {\bf R}  \end{math} is 
\begin{equation}
\psi =\psi ({\bf r}-{\bf R}),
\label{Ciuchi2}
\end{equation}
\begin{math} v \end{math} is the relative velocity and 
\begin{math} \sigma  \end{math} is the \begin{math} e^- p^+ \end{math} 
cross section.

The relative velocity value employed by Ciuchi {\it et al} is copied from 
our paper on exploding wires thus arriving at a theory of exploding 
batteries\cite{explode}. Absurdities would also arise from Ciuchi {\it et al} 
taking our numbers from a paper describing neutron rates in lightening bolts. 
All these papers of ours are cited and numbers copied from them even though they 
are clearly irrelevant for describing neutron production on metal hydride 
cathodes.  

\section{Many Body Wave Functions \label{mbw}} 

The wave function problem not properly taken into account by Ciuchi {\it et al} 
is that  the time honored Eqs.(\ref{Ciuchi1}) and (\ref{Ciuchi2}) hold 
true if and only if there is precisely one electron and one proton in the 
initial incoming quantum state. If one is trying to treat 
\begin{math} N \end{math} protons and \begin{math} N \end{math} electrons 
then the charge neutral wave function Eq.(\ref{Ciuchi2}) would have to be 
replaced by 
\begin{equation}
\Psi=\Psi({\bf r}_1,{\bf r}_2,\ldots,{\bf r}_N,{\bf R}_1,{\bf R}_2,\ldots,{\bf R}_N)
\label{mbw1}
\end{equation}  
with spins and other degrees of freedom left implicit. 
Thus, for (say) \begin{math} N\sim 10^{16}  \end{math} 
participating in a surface plasmon, the probability \begin{math} |\psi(0)|^2 \end{math} 
employed by Ciuchi {\it et al} does not in reality exist. The many body version of the 
probability of finding an electron on top of a proton is described by the correlation 
function  
\begin{equation}
C= \frac{1}{N} \left<\Psi \right| \sum_{i=1}^N \sum_{j=1}^N 
\delta({\bf r}_i-{\bf R}_j)\left|\Psi \right>
\label{mbw2}
\end{equation}  
or the quantum field theory equivalent. 
What is here crucial is that the cathode is hot. It is sufficient;y hot for the cathode to glow optically 
and light up the laboratory. Thus one must employ a thermal average 
\begin{equation}
C_T= \frac{1}{N} \left< \sum_{i=1}^N \sum_{j=1}^N 
\delta({\bf r}_i-{\bf R}_j)\right>_T
\label{mbw3}
\end{equation}  
at an optical noise temperature that we have theoretically estimated\cite{WL2:2007} 
to be \begin{math} T\sim 5000 K^{\rm o}  \end{math} in agreement with experiment\cite{Cirillo:2012}.
As one must, we employ 
\begin{math} C_T \end{math} and {\em not} \begin{math} |\psi (0)|^2 \end{math} 
for the plasma physics problem at hand. It is this truncation from the many body collective 
aspect [\begin{math}C_T\end{math}] to the two body 
[\begin{math}|\psi(0)|^2\end{math}] which is at the heart of the difference 
in their and our estimate of the rates. The plasmon modes contributing to 
Eq.(\ref{mbw3}) determine the parameter \begin{math} \beta \end{math} as shown in 
our work\cite{WL1:2006,WL2:2007} on metal hydride cathodes. 

\section{Concluding Statement \label{conc}}

No significant argument has been provided against our nuclear physics results. 
The experimental evidence of neutron production and nuclear transmutations in
properly designed plasma discharge electrolytic cells\cite{Cirillo:2012} agrees with 
our theoretical analysis and belies the theoretical arguments given in\cite{Ciuchi:2012}
against a hefty production of neutrons in hydride cells.


\begin{thebibliography}{12}

\bibitem{Ciuchi:2012}
S. Ciuchi, L. Maiani, A. D. Polosa, V. Riquer, G. Ruocco and M Vignati,
arXiv:1209.6501v1 [nucl-th] 28 Sep 2012.

\bibitem{WL1:2006}
A. Widom and L. Larsen, {\it Eur. Phys. J.} {\bf C 46}, 107 (2006).

\bibitem{WL2:2007}
A. Widom and L. Larsen, 
arXiv:0608059v2 [nucl-th] 25 Sep 2007. 

\bibitem{Polosa} 
A. D. Polosa and M. Vignati, Roma I Internal Report [June 4, 2012].

\bibitem{Cirillo:2012}
D. Cirillo, R. Germano, V. Tontodonato, A. Widom, Y.N. Srivastava, E. Del Giudice, 
and G. Vitiello {\it Key Engineering Materials} {\bf 495}, 104 (2012). 

\bibitem{explode}
If compared with the Naples neutron production experiment, 
then the battery would have exploded and perhaps knocked out the 
laboratory or the first floor of the building. 


\end{thebibliography}
\end{document}